\documentclass[onecolumn,amsmath,amssymb, 12pt]{revtex4}
\usepackage{graphicx}% Include figure files
\usepackage{dcolumn}% Align table columns on decimal point%
\usepackage{bm}% bold math
\begin{document}
\title{The fractal/small-world dichotomy in real-world networks}
\author{G\'abor Cs\'anyi}\email{gabor@csanyi.net}
\affiliation{TCM Group, Cavendish Laboratory, University of Cambridge \\Madingley Road, Cambridge CB3 0HE, United Kingdom}
\author{Bal\'azs Szendr\H oi\footnote{Corresponding author}}
\email{szendroi@math.uu.nl}
\affiliation{Department of Mathematics, Utrecht University \\ 
PO. Box 80010,NL-3508 TA Utrecht, The Netherlands}
\date{\today}%

\begin{abstract}
We draw attention to a clear dichotomy between small-world networks
exhibiting exponential neighborhood growth, and fractal-like networks
where neighborhoods grow according to a power law. This distinction
is observed in a number of real-world networks, and is related to the
degree correlations and geographical constraints. We conclude by
pointing out that the status of human social networks in this
dichotomy is far from clear.

\end{abstract}
\maketitle

\section{Introduction} 

The idea of small-world networks~\cite{pool, milgram}, large systems which one can 
traverse in a few steps, has become extremely wide-spread in modern-day 
scientific and popular thinking. Many networks have been classified as 
``small worlds''~\cite{watts_strog}, ranging from social acquaintance networks 
through technological ones to networks in biology.

More recently, the notion of ``small world'' has been given a precise
meaning: a network is a small world, if the average distance between
nodes is at most a logarithm of total system
size~\cite{newman_review}.  This scaling behavior is one of the basic
properties of random graphs in the sense of Erd\H os and
R\'enyi~\cite{er}, though these latter networks are not ``small
worlds'' in the more restrictive sense of~\cite{watts_strog}, since
they have low clustering.  Focusing our attention to distances in
networks, it is intuitively obvious that average distances depend on
the quantitative growth of vertex neighborhoods; in particular,
logarithmic average distance corresponds to exponential growth in the
size of neighborhoods (precise definitions are given in the next
section). Conversely, in networks where neighborhoods of nodes grow
according to a power law rather than exponentially, average distances
also grow as a power of system size rather than its logarithm. Hence
such networks are not small worlds in the technical sense of the
term. Since a power law growth in neighborhood size is a discrete
analogue of fractal growth~\cite{mandel, vicsek}, we call this the
{\it fractal/small-world dichotomy}.

The main point of the paper is to demonstrate that this dichotomy can be clearly 
observed in classes of real-world networks. Social networks such as scientific
collaboration networks, and the Internet at router level are typical examples of 
small worlds in the strict sense. On the other hand, 
networks with strong geographical constraints, such as power grids or 
transport networks, are examples of networks with fractal scaling. 
It may not be surprising that the topology of these 
networks is strongly constrained by their geographical embedding; however, 
one example, the power grid, was consistently classified as a 
``small--world network'' so far~\cite{watts_strog, watts, classes, ba_review, 
newman_review}. Fractal scaling sheds new light on the effect of
long-range connections, different from the original interpretation of 
Watts and Strogatz~\cite{watts_strog} and more in line with the theoretical ideas
of~\cite{euclidean}: in such networks, even long-range connections are
constrained by Euclidean distance, and hence cannot
give rise to true small-world behavior. 

\section{Scaling of neighborhoods} 

Begin with the case of a graph $G$ on an infinite set of nodes, 
with every node connected to finitely many others only. Fix a node $v\in G$, 
and denote by $N_v(r)$ the size of the {\em radius $r$ neighborhood} of $v$, 
the number of nodes of $G$ which can be reached from $v$ in at most $r$ steps. 
Consider the following two limits, which may or may not exist:
\begin{equation}\label{ddef} d = \lim_{r\to\infty} \frac{\log N_v(r)}{\log r}\end{equation}
and
\begin{equation}\label{alphadef} \alpha = \lim_{r\to\infty} \frac{\log N_v(r)}{r}.\end{equation}
Clearly if a finite, nonzero limit for $d$ exists, then $\alpha=\infty$; 
conversely, if $\alpha$ is finite then $d=0$. It is easy to see that for connected 
$G$, if either of the limits exists then it is independent of $v$. It is also 
immediate that a finite and nonzero value for the limit corresponds in the two 
cases to the following scaling laws: 
\begin{equation}\label{dscale} N_v(r) \sim r^d\end{equation}
and
\begin{equation}\label{alphascale}  N_v(r) \sim e^{\alpha r}\end{equation}
respectively, both valid for large $r$. Equation~(\ref{dscale}) is the discrete 
analogue of {\it fractal scaling}~\cite{mandel,vicsek}, 
with $d$ corresponding to the mass dimension of a fractal.

Our main interest is in finite networks $G$, where a scaling law can only hold in 
some finite range. As shown by several examples below, 
the size of neighborhoods is often well approximated for $1<r<L$ by a uniform 
scaling law, either~(\ref{dscale}) or~(\ref{alphascale}), and at most a constant 
proportion of nodes lies outside this range. Under this assumption, the total 
number~$N$ of nodes of~$G$ is
\begin{equation}\label{dtot} N \sim \sum_{r=1}^L r^d \sim L^d \end{equation}
and
\begin{equation}\label{alphatot}  N\sim  \sum_{r=1}^L e^{\alpha r} \sim e^{\alpha L}\end{equation}
respectively. On the other hand, let $l_v$ be the average distance between $v$ 
and other nodes in the graph~\cite{watts_strog}. This quantity can then be 
expressed as
\begin{equation}\label{dave} l_v \sim \frac{1}{N}\sum_{r=1}^L r\cdot r^d \sim \frac{L^{d+1}}{N} \end{equation}
and
\begin{equation}\label{alphaave} l_v \sim \frac{1}{N}\sum_{r=1}^L r\cdot e^{\alpha r} \sim \frac{L\cdot e^{\alpha L}}{N}\end{equation}
in the two cases. By equations~(\ref{dtot})--(\ref{dave}), 
respectively~(\ref{alphatot})--(\ref{alphaave}), $l_v$ and $L$ are 
proportional, independent of system size; and more importantly, fractal 
scaling~(\ref{dscale}) implies
\begin{equation}\label{ddiam} l_v \sim N^{1/d},\end{equation}
whereas the exponential scaling law~(\ref{alphascale}) leads to 
\begin{equation}\label{alphadiam} l_v \sim \log N.\end{equation} 
Logarithmic scaling of average distance with system size is nowadays taken as the 
definition of {\it small-world} behavior~\cite{newman_review}. In the original 
use of the term~\cite{milgram, watts_strog}, a small-world network was one with 
``surprisingly small'' average distance compared to its size. Logarithmic scaling 
of average distance with size is a precise way to characterize the small-world 
property in growing network processes, where there is a meaningful range of 
system sizes. For a fixed network $G$, this is still not sufficient; 
however, the scaling law~(\ref{alphascale}) is meaningful. Hence~(\ref{alphascale}) 
will be called {\it small-world scaling}. A slight extension is in 
fact necessary, since some networks are known to have smaller than logarithmic 
average distances~\cite{ultrasmall, chunglu}. These networks must have a 
faster-than-exponential neighborhood growth for some range of radii. 
Following~\cite{newman_review}, such networks will also be referred to 
as small worlds in this paper. 

Expressions~(\ref{dscale})--(\ref{ddiam}) 
versus~(\ref{alphascale})--(\ref{alphadiam}) 
(and generalizations to super-exponential behavior) present a dichotomy. 
On the one hand, there are graphs with fractal-like growth 
of neighborhoods, coupled to a power-law 
diameter. On the other hand, there are graphs with small-world 
behavior, characterized by at least exponential neighborhood growth coupled 
to at most logarithmic diameter. 

\section{Real-world networks} 

\subsection{Small-world scaling} 

\begin{figure}
$\begin{array}{c@{\hspace{.34in}}c}
\scalebox{0.39}{\includegraphics{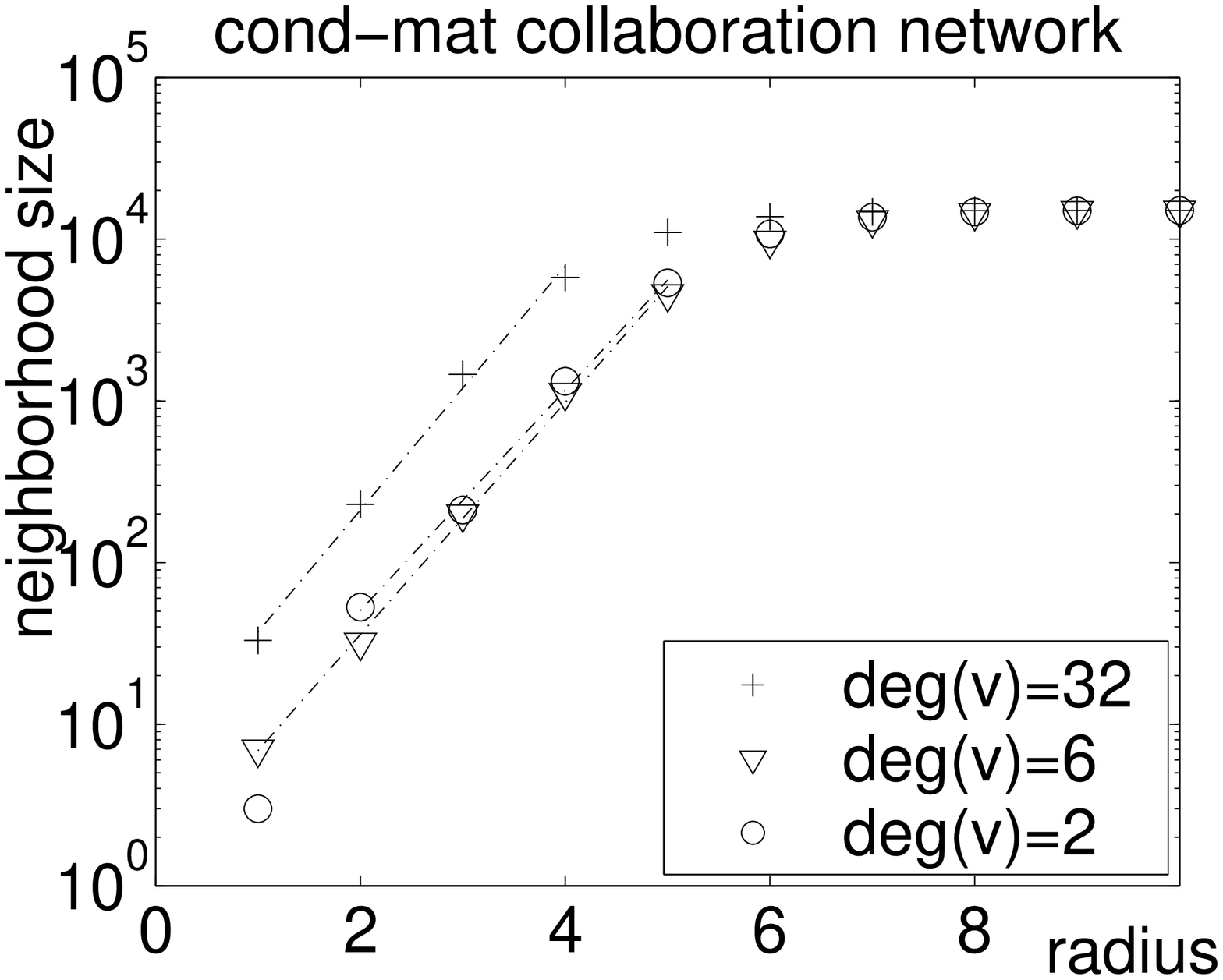}}&
\scalebox{0.39}{\includegraphics{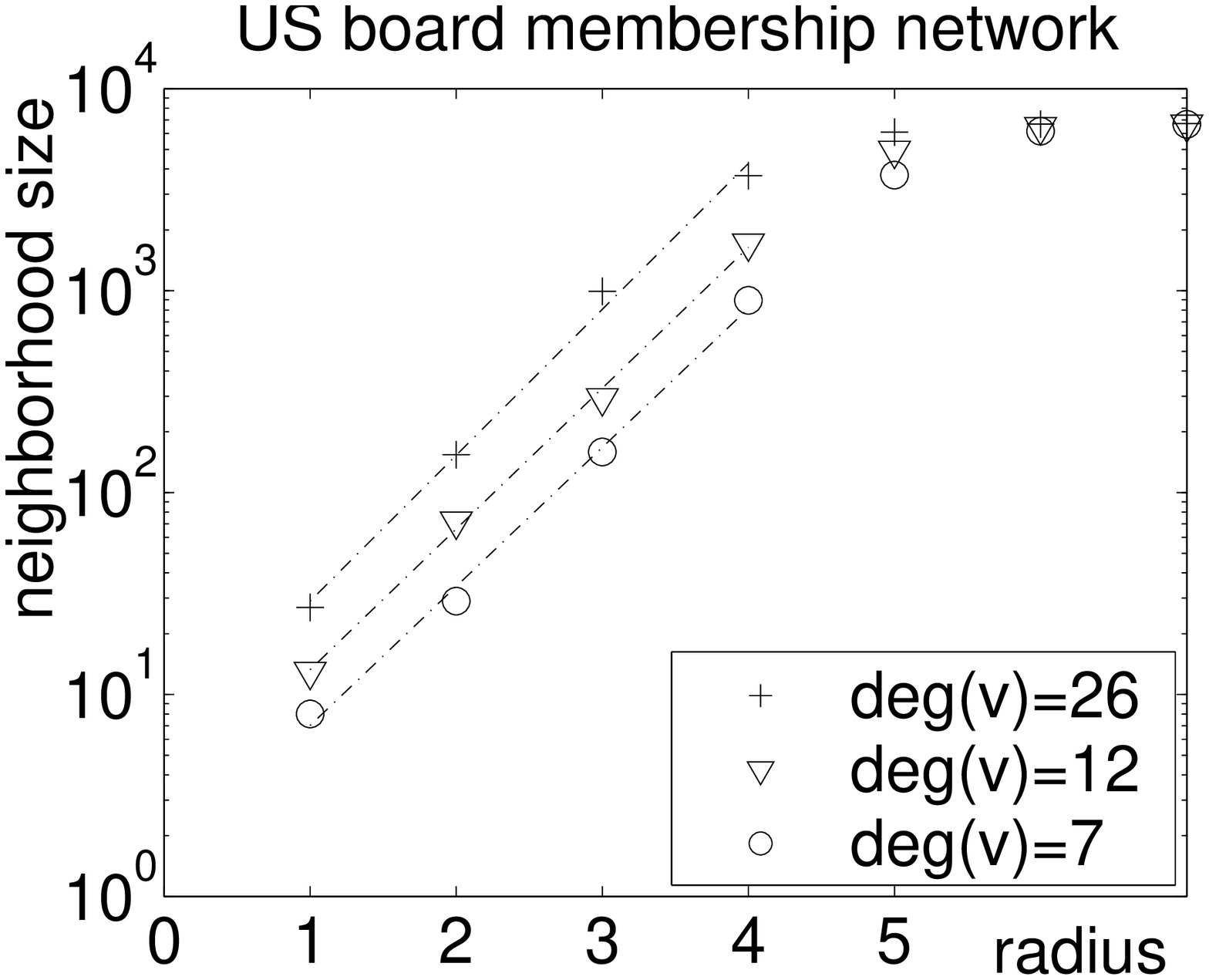}}
\end{array}$
\caption{\it The scaling of neighborhood size for typical vertices in two 
small-world social networks. The cond-mat collaboration network~\cite{new_pnas} has 
$N=17636$ nodes and $E=55270$ edges. The 1999
US board membership network~\cite{boards} 
includes data about 916 large US companies and a total of $N=7680$ board members 
connected by $E=55436$ links.}
\label{condmatfig}
\end{figure}

We proceed to show that this fractal/small-world dichotomy is actually detectable 
in a variety of real-world networks, many of which have loosely been classified so 
far as ``small worlds''. The issue is complicated by the fact that, because of 
saturation effects, (super)exponential scaling is only detectable in networks 
which are really ``large''; for a network of average diameter $l\ll10$, there 
will only be two or three meaningful data points on the $r-N_v(r)$ plot, and 
any statement that the points really represent exponential growth may be tenuous. 
For a case in evidence, consider Figure~\ref{condmatfig}, where we plotted the 
growth of $N_v(r)$ against~$r$ for typical nodes $v$ in two social networks: 
the scientific collaboration network corresponding to the corned-mat preprint 
archive~\cite{new_pnas}, and the network of board members of large US companies
in 1999, with links between people sitting on at least one board 
together~\cite{boards}. The data is compatible with exponential growth in both 
cases, and the scaling exponents are in good agreement. This supports the 
conclusion that these social networks satisfy small-world scaling, but one 
would obviously wish to see a few more data points.  

More generally, there is ample indirect evidence for the existence of networks with 
small-world scaling. As shown by~\cite{ajb, faloutsos} and consequently by many 
other groups, real-world networks often have a power-law tail in their degree 
distribution. Power-law distribution can be generated by preferential 
attachment~\cite{ba, boll_spen}, and indeed the model~\cite{ba} builds a small 
world~\cite{boll_rior}. It was consequently shown in~\cite{ultrasmall, chunglu} 
that the diameter of random scale-free networks is at most logarithmic, as long 
as there is no correlation between the degrees of neighboring vertices;  
a positive correlation between vertex degrees is expected to decrease 
average distances even further. 
Hence real-world graphs with a power law distribution and 
positive or no correlation between degrees are small worlds, exhibiting
exponential scaling once system size is sufficiently large.  

One example that merits further discussion is the Internet graph, which one 
treats separately at the inter-domain (AS) and at the router 
levels. The degree sequence of both levels was shown to possess a power law tail 
by~\cite{faloutsos}, confirmed also by later measurements~\cite{govindan}. On the 
other hand, both these networks were claimed~\cite{faloutsos} to possess fractal 
scaling~(\ref{dscale}) in neighborhood growth. Interestingly, the correlation 
between degrees of neighboring vertices is quite different in 
the two cases. In the AS level network, degrees are negatively 
correlated~\cite{vazvespinternet}, which was claimed to be a generic feature of 
technological networks~\cite{newman_corr}. At router level however, there is a 
slight positive correlation~\cite[Fig.~1d]{vazquez}. 

\begin{figure}
\scalebox{0.4}{\includegraphics{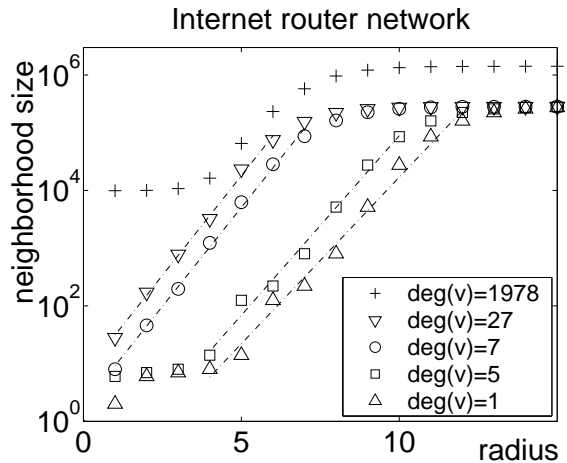}}
\caption{\it Neighborhood scaling for typical vertices in the Internet router 
network~\cite{govindan, routerdata}, with $N\approx 3\times 10^5$. The topmost 
graph was moved vertically for 
better visibility. For small degree vertices, exponential scaling is apparent; 
for a vertex with large degree, saturation is very strong.}
\label{routerfig}
\end{figure}

Turning to the question of scaling, the AS level network, with about $10^4$ 
vertices, is too small; 
the neighborhood growth plots are inconclusive, though consistent with 
exponential scaling. Indeed, despite its negative degree correlations, 
one expects the AS level network to be a small world, since there are no physical 
restrictions on the placement of links. 
The router network is constrained by geography to some extent, and has 
degree-independent clustering coefficients~\cite{rav_barab}, thought to be a 
characteristic of geographic networks. On the other hand, its power law and 
global topology are driven partially by preferential attachment~\cite{topo}. 
The latter effect is strong enough to create a small world~\cite{as}:
its exponential scaling is depicted on Figure~\ref{routerfig}.

One point to note is that a power law tail in the degree sequence of a network
does not necessarily imply that the graph is a small world. 
The model of~\cite{euclidean} 
embeds a power law graph in a Euclidean lattice, and for some range of parameters, 
fractal scaling survives. However, this model, as well as the model 
of~\cite{vazvespmodel} with similar properties, have strong negative correlation 
between vertex degrees, to our knowledge not observed in real-world networks.  

\subsection{Fractal scaling} 

We turn to real-world networks with 
fractal scaling, our examples coming from the class of
geographical networks. It was noted before 
that geographic networks behave differently 
from typical small worlds such as the World Wide Web and collaboration networks 
in several respects; their degree distribution need not follow a power 
law~\cite{classes} and they appear to have trivial degree-clustering 
correlation~\cite{rav_barab}. 

\begin{figure}
$\begin{array}{c}
\begin{array}{c@{\hspace{.32in}}c}
\scalebox{0.39}{\includegraphics{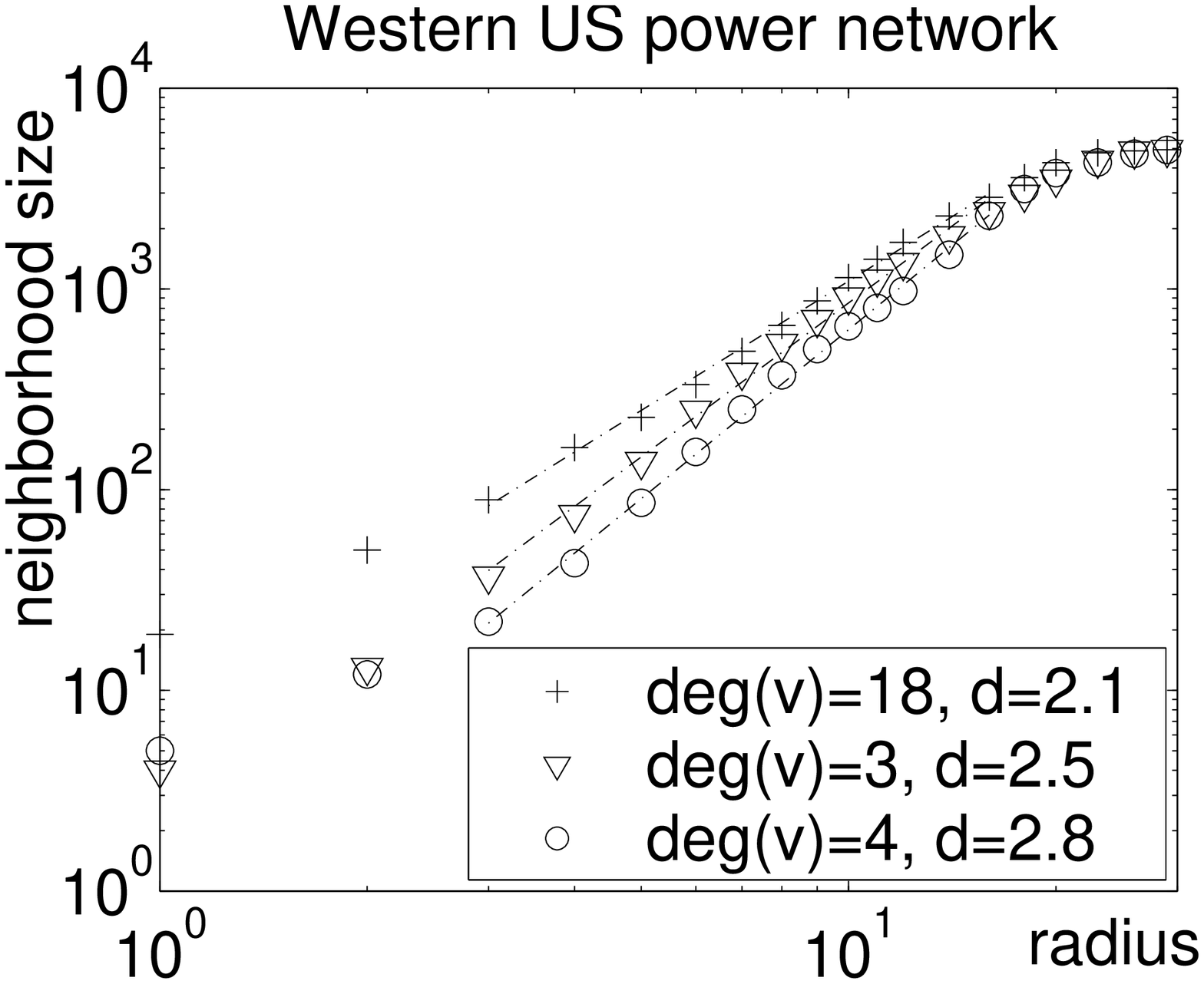}} &
\scalebox{0.39}{\includegraphics{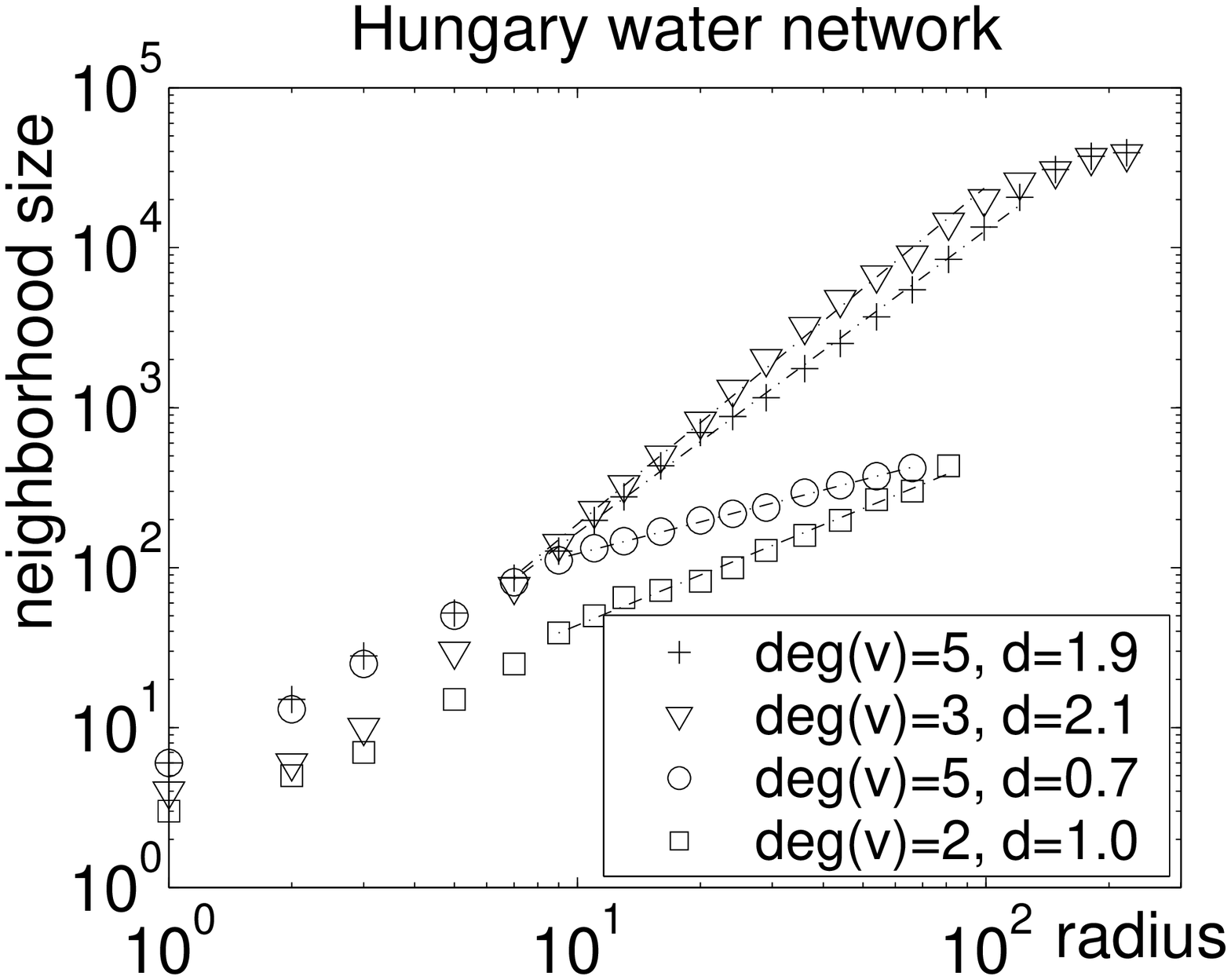}} \end{array}\\
\scalebox{0.4}{\includegraphics{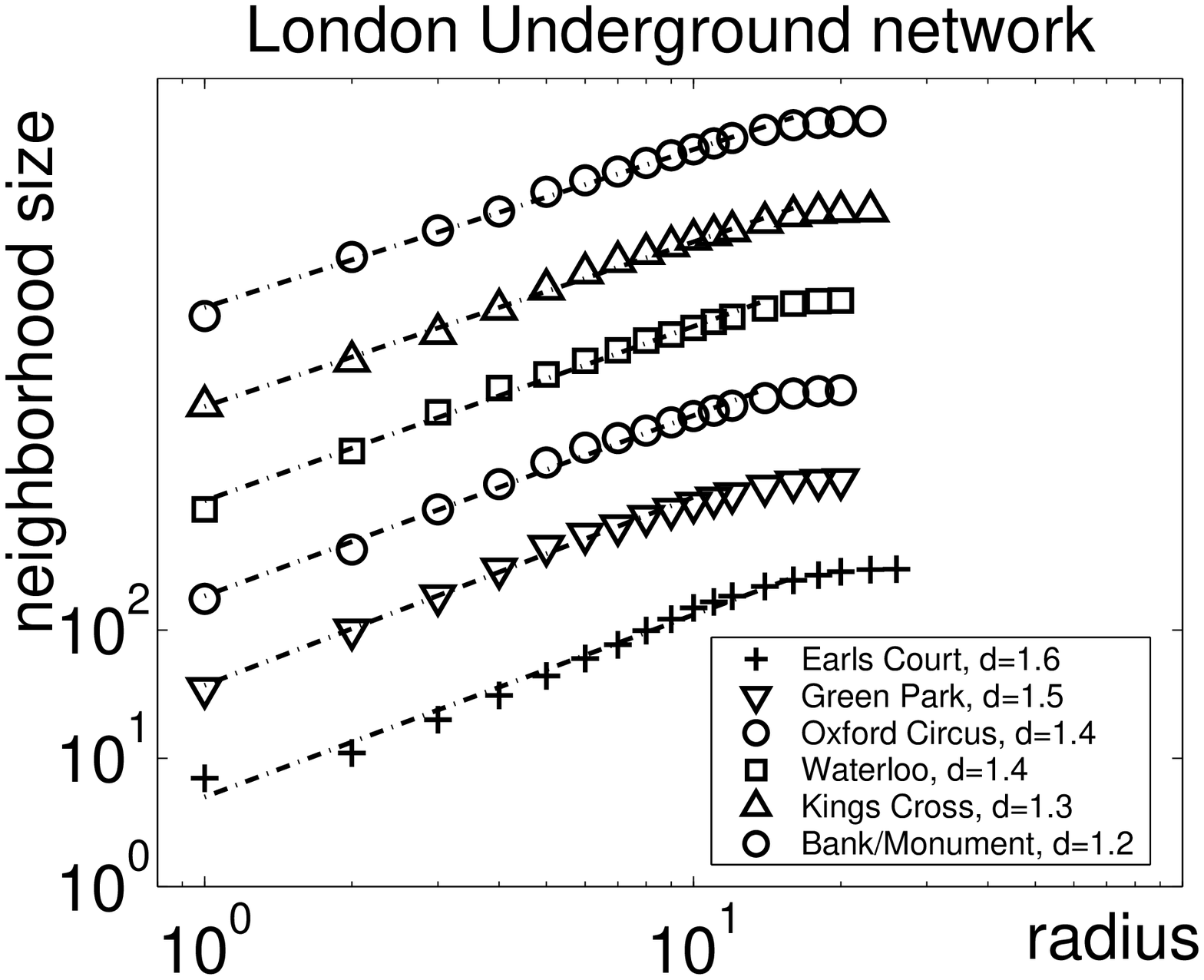}}
\end{array}$
\caption{\it Scaling of neighborhoods for some geographical networks. 
Scaling for the power grid~\cite{watts_strog}, with $N=4941$ nodes and $E=6594$ 
edges, is illustrated on the left panel. The fractal exponents lie between $2$ and 
$3$. The water network of Hungary has $N=41495$ nodes and $E=50252$ 
edges. The first two vertices of the right panel belong to the giant component of 
size $N_1=39247$ and fractal dimension $d\approx 2$. The second pair of vertices 
belongs to the second largest component containing $N_2=465$ vertices, and 
exhibits fractal scaling with exponents around or below 1. 
The London Underground network has $N=300$ nodes connected by $E=348$ edges; 
the scaling graphs from different starting stations are shifted vertically 
for better visibility.}
\label{powerfig}
\end{figure}

The power grid of the Western US is a much studied example, appearing
already in~\cite{watts_strog}. As the left panel in
Figure~\ref{powerfig} shows, it satisfies fractal
scaling~(\ref{dscale}) with exponents \cite{exponents} lying between 2 and 3. Hence
networks structurally equivalent to the US power grid have a larger
than logarithmic diameter; under the strict definition, the power grid
is not a small--world network.

As further examples, consider two other geographical networks. The water network of 
Hungary, with major water distribution centers as nodes connected by water pumping 
lines as edges, is studied on the right panel of Figure~\ref{powerfig}. 
A typical example of a 
transport network, the London Underground, is investigated on the lower panel of
Figure~\ref{powerfig}. Stations are represented by nodes,  
and two nodes are connected by an edge if the corresponding stations are neighbors 
on some Underground line (including Thameslink). 
Both networks satisfy fractal scaling~(\ref{dscale}). 

Note that the power grid, water and Underground networks are all embedded in a 
$d=2$ dimensional space, the surface of the Earth. For the giant 
component of the water network, we indeed obtain scaling exponents 
$D\approx 2$, which seems to indicate that the distribution of nodes and edges 
follows the geographical constraints. For the power grid, some of the obtained 
exponents are significantly higher than $2$. The reason for this is the existence 
of long-range connections, already discussed by Watts and Strogatz 
in~\cite{watts_strog, watts}. As we see here, long-range power supply lines have a 
significant effect on the measured fractal dimension, but they are not sufficient 
to turn the power grid into a small--world network. Thus the long-range connections
are not distributed randomly, as anticipated by~\cite{watts_strog}, but they too
respect the Euclidean structure. This is more in line with the theoretical 
discussion of~\cite{mok, euclidean}, where long-range connections are introduced 
with a probability that depends on Euclidean distance.  
In the case of the London Underground network, the fractal exponents are
strictly between $1$ and $2$, indicating the fact that the Underground network 
penetrates only a (fractal) subset of the $d=2$ dimensional surface of Greater 
London, which however is strictly larger than a $d\approx 1$ dimensional set that 
a few isolated (linear) underground lines could cover.

\subsection{Mixed scaling} 

For completeness, we briefly discuss the possibility of mixed behavior: 
the case of a network exhibiting fractal and small-world scalings at different
length scales. As discussed in~\cite{watts_newman}, the original model of
Watts and Strogatz~\cite{watts_strog} exhibits this behavior for small values
of the rewiring parameter: at small scales, the network retains its Euclidean 
structure, but at large scales it is a small world. The opposite behavior is 
perhaps more natural in social networks. Consider a network obtained by 
placing a small-world network of size $N>>0$ in every lattice point of a lattice 
$\Lambda$, and connecting vertices belonging to different lattice points with some 
fixed probability if the corresponding lattice points are adjacent in the lattice. 
The resulting network is a simple model of a collection of cities with
small-world populations, where people only socialize with others in their own
or in neighboring cities. For neighborhood sizes $N_v(r)<<N$, this network 
exhibits small world scaling, whereas on large scales the underlying lattice
dominates and the scaling becomes fractal. 

For real-world networks, we have not found conclusive evidence for this kind 
of behavior, because a network must indeed be very large to show such features. 
Note that, as discussed above, exponential scaling in itself is already difficult
to demonstrate unless the network is sufficiently large.

\subsection{Relationship to other network measures}

\begin{figure}
\scalebox{0.6}{\includegraphics{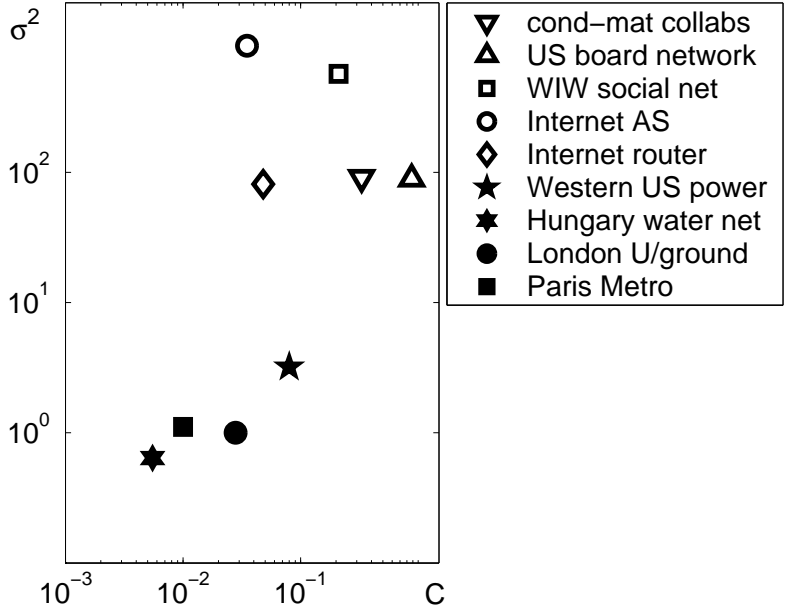}}
\caption{\it The average local clustering $C$ and degree variance $\sigma^2$ for some
small-world networks (empty signs) and fractal networks (filled signs)}
\label{scatterplot}
\end{figure}

As we discussed above, the small-world property in real-world networks is typically
associated with a strongly right-skewed degree distribution, such as a power law. 
On the other hand, as discussed extensively by~\cite{watts_strog}, small worlds
also contain many triangles. In Figure~\ref{scatterplot}, 
we plot the average local clustering $C$, a local measure of triangle density, 
and the degree distribution variance $\sigma^2$ for some networks.  
We observe a separation into two clusters, with small worlds characterised by large $C$
and $\sigma^2$ values, and fractal networks typically having smaller values.  
Apart from all the networks appearing in our earlier discussion, we included some
additional networks, such as the Paris Metro fractal network, and the small-world 
web-based social network WIW~\cite{csz}.

\section{Conclusion}

We have demonstrated a clear dichotomy between large real-world 
networks which are small worlds with exponential neighborhood growth, 
and fractal networks with a power law growth. Typical examples of the former are 
networks with little or no geographical confinement, 
such as collaboration networks and the Internet. 
The latter are typified by networks strongly constrained by geography. 
It also emerged that in the latter case, the fractal exponents vary
considerably; so instead of ``hop-plot'' diagrams averaged over the whole
network, it is preferable to study the scaling of neighborhood size
$N_v(r)$ with radius $r$ for individual vertices $v$.

One question that emerges from our discussion is whether the social 
network of humans~\cite{pool, milgram, watts_new} is a small world 
in the strict sense of neighborhood growth. The two examples of social networks
studied in this paper do form small worlds indeed, even though geographical 
proximity obviously plays some role in their formation (this is discussed
explicitly in~\cite{boards} for the US board membership network). However, 
here Kleinfeld's argument~\cite{kleinfeld} definitely applies: in these small 
worlds, the majority of actors belong to an extremely homogeneous population 
(Western scientists, US entrepreneurs) mostly on one side of racial and 
class barriers, united by common profession. 

On a global scale, the answer is much less clear. Is human society really strongly
connected, with sufficiently many non-local links to lead to exponential 
growth in the number of acquaintances? Or are contacts on a large scale
restricted by geographical position as well as different social barriers, 
so that one only has circles of acquaintances growing in size according to a 
power law? The importance of this issue is emphasized by the fact that, 
the frivolous example of gossip aside, social contact networks are involved not only
in the spread of advertising and other essential information, but also that 
of viruses, for example in the case of sexually transmitted diseases~\cite{klov}.
We believe that the fractal/small-world dichotomy is central to the true understanding
of the structure of massive real-world graphs.

\section*{Acknowledgments} 
We thank Gerald Davis, Mark Newman and Duncan Watts for providing network data.


\begin{thebibliography}{99}
\bibitem{exponents}The range of the fractal scaling approximation
$N_v(r)\sim r^d$ was determined using the following procedure: let the
average percentage error of the best log--linear approximation on the
radius interval $(r,R)$ be $\langle E(r,R)\rangle$.  Keeping one end fixed
and treating the other as variable results in an average error function
$\langle E(r,R)\rangle$ which is uniformly small up to a point where it
undergoes a sudden increase. This gives the choice of reasonable upper and
lower bounds for the validity of the fractal scaling approximation.
\end{thebibliography}
\end{document}